\documentclass{Interspeech}

\interspeechcameraready

\title{Delayed-KD: Delayed Knowledge Distillation based CTC for Low-Latency Streaming ASR}

\author[affiliation={1}]{Longhao}{Li}
\author[affiliation={1}]{Yangze}{Li}
\author[affiliation={1}]{Hongfei}{Xue}
\author[affiliation={2}]{Jie}{Liu}
\author[affiliation={2}]{Shuai}{Fang}
\author[affiliation={2}]{Kai}{Wang}
\author[affiliation={1}{*}]{Lei}{Xie}

\affiliation{Audio, Speech and Language Processing Group (ASLP@NPU)}{\\Northwestern Polytechnical University}{China}
\affiliation{}{Huawei Cloud}{China}
\email{lhli@mail.nwpu.edu.cn, lxie@nwpu.edu.cn\thanks{*Corresponding author.}}

\usepackage{comment}
\usepackage{multirow}
\usepackage{arydshln}

\begin{document}

\maketitle

\begin{abstract}
    
    CTC-based streaming ASR has gained significant attention in real-world applications but faces two main challenges: accuracy degradation in small chunks and token emission latency. To mitigate these challenges, we propose Delayed-KD, which applies delayed knowledge distillation on CTC posterior probabilities from a non-streaming to a streaming model. Specifically, with a tiny chunk size, we introduce a Temporal Alignment Buffer (TAB) that defines a relative delay range compared to the non-streaming teacher model to align CTC outputs and mitigate non-blank token mismatches. Additionally, TAB enables fine-grained control over token emission delay. Experiments on 178-hour AISHELL-1 and 10,000-hour WenetSpeech Mandarin datasets show consistent superiority of Delayed-KD. Impressively, Delayed-KD at 40 ms latency achieves a lower character error rate (CER) of 5.42\% on AISHELL-1, comparable to the competitive U2++ model running at 320 ms latency.
\end{abstract}

\section{Introduction}

Streaming automatic speech recognition (ASR) has attracted significant attention in real-world applications, aiming to ensure recognition accuracy with low latency. Currently, the predominant end-to-end ASR models include Connectionist Temporal Classification (CTC)~\cite{graves2006connectionist}, Recurrent Neural Network Transducer (RNN-T)~\cite{graves2012sequence} and Attention-based Encoder-Decoder (AED)~\cite{chan2015listen, chorowski2015attention}. To facilitate streaming capabilities in the encoder of these end-to-end models, chunk-based methodologies are typically employed as a simple-yet-effective approach. Particularly, CTC is often favored for integration due to its architectural simplicity.

However, CTC-based streaming ASR faces two main challenges: accuracy degradation in small chunks~\cite{tang2023reducing} due to limited context~\cite{kumar2024xlsr} and the inherent non-blank token emission latency issue~\cite{fang2024mamba}, where CTC spikes lag behind non-streaming models. To address the issue of accuracy degradation in small chunks, several methods have been proposed to mitigate chunk-induced losses. For instance, Fast-U2++ \cite{liang2023fast} employs a hybrid chunk strategy, utilizing small chunks in lower encoder layers for early partial result output and large chunks in upper layers to compensate for performance degradation. USIDE-T~\cite{zhao2024cuside} introduces SimuNet, a history-frame-based simulation network, to approximate full-context information and enable smaller chunk sizes. While these strategies partially offset the performance decline caused by smaller chunks, they still underperform under tiny chunk conditions. To address the non-blank token emission latency issue, techniques like FastEmit~\cite{yu2021fastemit}, Peak-first CTC~\cite{tian2023peak}, and Delay-Penalty \cite{kang2023delay} encourage early non-blank token emission in the loss function. Although these methods effectively reduce CTC spike lag, they still suffer from significant model latency. Since token emission occurs at the end time of the chunk, this means that while the CTC spike lag is reduced, the token emission lag remains unaddressed. Additionally, these techniques often offer limited accuracy improvements. 

Streaming ASR often suffers from a performance gap compared to non-streaming ASR due to limited context~\cite{kojima2021knowledge}. Knowledge distillation~\cite{hinton2015distilling, shim2023knowledge, seth2024stable} has demonstrated its effectiveness in reducing this gap and enabling early token emission~\cite{liang2023fast}. However, the inherent token emission delay in streaming CTC training limits the effectiveness of direct frame-by-frame distillation~\cite{yu2020dual, inaguma2021alignment}. Previous studies~\cite{kurata2019guiding, tian2022bayes, kim2024guiding} have primarily focused on explicit forced alignment with CTC for distillation but overlook the information discrepancy between streaming and non-streaming models, leading to limited effectiveness of the forced alignment learning. This highlights the need for a low-latency and efficient CTC alignment distillation method that can effectively balance accuracy and latency.

In this paper, we propose Delayed-KD to improve recognition accuracy in low-latency streaming scenarios. Delayed-KD introduces a Temporal Alignment Buffer (TAB) with a tiny chunk size during training, enabling delayed knowledge distillation of CTC posterior probabilities from a non-streaming teacher model to a streaming student model. Specifically, TAB introduces a controlled delay range, allowing distillation within this range to select the chunk with minimal KL divergence loss, effectively overcoming the limitations of frame-by-frame distillation. Experimental results on the AISHELL-1~\cite{bu2017aishell} Mandarin dataset demonstrate that Delayed-KD achieves a 9.4\% relative CER reduction compared to another competitive solution U2++~\cite{wu2021u2++} at 40 ms latency. Notably, Delayed-KD at 40 ms latency achieves comparable performance to U2++ at 320 ms latency. Furthermore, Delayed-KD demonstrates superior low-latency and high-accuracy capabilities compared to other advanced streaming models. On the large-scale, multi-domain Mandarin WenetSpeech~\cite{zhang2022wenetspeech} dataset, in rescoring mode at 40 ms latency, Delayed-KD achieves a 27.9\% and 33.1\% relative CER reduction on \textit{Test\_Meeting} set and \textit{Test\_Net} set, respectively. Additionally, TAB adjustment allows fine-grained control over the trade-off between token emission delay and accuracy.

\section{Method}

\subsection{Model architecture}

\begin{figure*}[t]
  \centering
  \includegraphics[width=\linewidth]{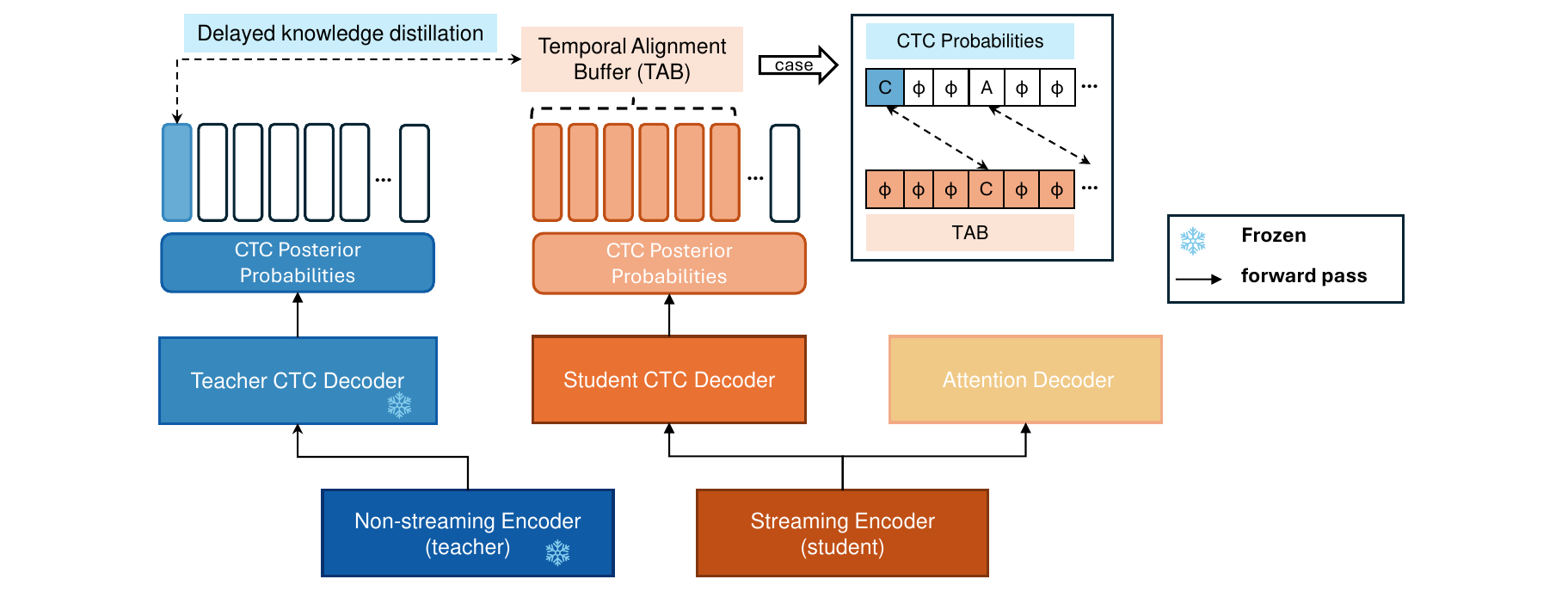}
  \caption{Model architecture of our proposed Delayed-KD}
  \label{fig:model_architecture}
\end{figure*}

As shown in Figure 1, the proposed model architecture, comprises three main components: the CTC branch of the non-streaming teacher model, the CTC branch of the streaming student model, and the attention branch of the streaming student model. Notably, the CTC branches of both the teacher and student models share an identical structure, each consisting of a shared encoder for modeling the context of acoustic features and a CTC decoder for aligning frames and tokens. The attention branch of the student model incorporates an Attention Decoder to model dependencies among tokens.

The shared encoder is constructed with multiple Conformer \cite{gulati2020conformer} layers, while the CTC decoder is composed of a linear layer followed by a log softmax layer. This design serves a dual purpose: firstly, it applies the CTC loss function to the softmax output during training, and secondly, it facilitates delayed knowledge distillation of frame-level CTC posterior probabilities between the teacher and student models. The Attention Decoder adopts the same architecture as traditional Transformer decoders, ensuring robust modeling of token dependencies.

\subsection{Training}

In this section, we mainly discuss two training details: delayed knowledge distillation and joint training. 

\subsubsection{Delayed Knowledge Distillation}

As shown in Figure~\ref{fig:model_architecture}, the fbank features of the audio are separately fed into the encoder of the non-streaming teacher model's CTC branch and the encoder of the streaming student model's CTC branch. This process generates two distinct frame-level CTC posterior probability distributions: one from the non-streaming model, which incorporates global context, and the other from the streaming model, which is limited to left-side context. 

During the distillation process, we employ Kullback-Leibler (KL) divergence as the distillation metric. To address the latency of CTC spikes in the streaming model compared to the non-streaming model, we introduce a \textbf{Temporal Alignment Buffer (TAB)} to mitigate misalignment issues during distillation. Specifically, during training, we adopt a tiny chunk size. TAB maintains a controlled delay relative to the non-streaming teacher model, enabling distillation to be performed within this delay range to select the chunk with the minimum KL divergence loss. The delay range is dynamically determined by a predefined range, which allows the selection of an optimal number of chunks to account for potential delays between the non-streaming teacher model and the streaming student model. Within this range, we calculate the KL divergence across various time delays based on chunks and select the delay that minimizes the loss as the final KL loss. By aligning the teacher model's CTC output with the student model's delayed CTC outputs (shifted right by a few chunks), we effectively mitigate mismatches in non-blank tokens during distillation. This approach significantly enhances the performance of the distillation process.

Based on the above process, we define the delayed knowledge distillation loss as follows:

\begin{align}
  L_{\text{distill}} &= \frac{1}{B T} \sum_{b=1}^{B} \sum_{t=1}^{T} \min_{\tau \in \mathcal{D}} \sum_{c=1}^{C} \text{KL}(p_{\text{student}}^{(b,t+\tau,c)} \parallel p_{\text{teacher}}^{(b,t,c)})
  \label{equation:distill_loss}
\end{align}
where $ B $ is the batch size, $ T $ is the number of time steps, $ C $ is the vocabulary size, and $ \mathcal{D} $ is the delayed knowledge distillation window range, which corresponds to the TAB. Specifically, $ \mathcal{D} = \{0, 1, 2, \dots, d\} $ defines the allowable delay range for aligning the student and teacher model outputs, where $ d $ is the maximum delay determined by the TAB size.

The KL divergence is used to align the CTC posterior probability distributions of the student and teacher models, and it is calculated as follows: 

\begin{align}
  \text{KL}(p_{\text{student}} \parallel p_{\text{teacher}}) &= \sum_{i} p_{\text{student}}(i) \cdot \log \frac{p_{\text{student}}(i)}{p_{\text{teacher}}(i)}
  \label{equation:kl_divergence}
\end{align}
where $p$ represents the posterior probability distribution of the CTC output from either the teacher or student model.

\subsubsection{Joint Training}

The overall loss function \( L_{\text{joint}} \) is a weighted combination of the ASR task loss \( L_{\text{asr}} \) and the delayed knowledge distillation loss \( L_{\text{distill}} \) as listed in the Equation~\ref{equation:distill_loss}, defined as:

\begin{align}
  L_{\text{joint}} &= L_{\text{asr}} + \alpha L_{\text{distill}}
  \label{equation:joint_loss}
\end{align}
where \( \alpha \) is a hyper-parameter that controls the contribution of the delayed knowledge distillation loss to the total loss. The ASR task loss \( L_{\text{asr}} \) is further decomposed into a weighted sum of the CTC loss \( L_{\text{CTC}} \) and the attention-based encoder-decoder (AED) loss \( L_{\text{AED}} \):

\begin{align}
  L_{\text{asr}}(\mathbf{x}, \mathbf{y}) &= \lambda L_{\text{CTC}}(\mathbf{x}, \mathbf{y}) + (1 - \lambda) L_{\text{AED}}(\mathbf{x}, \mathbf{y})
  \label{equation:asr_loss}
\end{align}
where \( \lambda \) balances the contributions of the CTC and AED losses, \( \mathbf{x} \) represents the input features, and \( \mathbf{y} \) denotes the target sequence.

\subsection{Decoding}

We use the same two-pass rescoring decoding strategy as in U2++~\cite{wu2021u2++}. In the first pass, the frame-synchronous CTC decoder generates a set of candidate hypotheses using prefix beam search. During the second pass, the attention decoder computes the scores for these n-best candidates. The final score for each candidate is then determined by combining these scores according to Equation~\ref{equation:s_final}. The candidate with the highest final score is selected as the optimal output. 

\begin{align}
  S_{\text{final}} &= \lambda\, S_{\text{CTC}}
  + S_{\text{AED}} 
  \label{equation:s_final}
\end{align}

\section{Experiments}

\subsection{Dataset}

We conduct experiments on two Mandarin Chinese datasets: AISHELL-1~\cite{bu2017aishell} (178 hours) and the large-scale, multi-domain WenetSpeech~\cite{zhang2022wenetspeech} (10,000 hours). For AISHELL-1, the test set contains 7,176 utterances, while WenetSpeech provides two test sets, \textit{Test\_Meeting} and \textit{Test\_Net}, which collectively contain approximately 33,100 utterances. 

\subsection{Experimental Setup}

All experiments in this study are implemented and evaluated within the WeNet toolkit~\cite{zhang2022wenet}. We use the standard Conformer U2++ model as our baseline and conduct experiments based on its framework. The parameters of the teacher non-streaming model are frozen during the training process. The hyper-parameters for both the encoder and decoder in Delayed-KD are consistent with those of U2++ to ensure a fair comparison. 

The input features are 80-dimensional fbank features, with a frameshift of 10 ms and a frame length of 25 ms. We also implemented SpecAugment and SpecSub in the same way as U2++. The modelling units consist of approximately 4,000 Chinese characters in the AISHELL-1 dataset and approximately 5,500 in the WenetSpeech experiments.

The label smoothing weight is configured to 0.01. For joint training with the cross-entropy loss, $\lambda$ is assigned a value of 0.3. The Adam optimizer, combined with a Warmup learning rate scheduler, is employed for all models. The initial learning rate is set to 0.001, and gradient clipping is applied with a threshold of 5.0. The learning rate undergoes a warm-up phase over the first 25k steps. The total training steps are 200k for AISHELL-1 and 130k for WenetSpeech. During the decoding phase, for AISHELL-1, we average the top 20 models based on the lowest dev loss, while for WenetSpeech, the top 10 models are averaged.

\subsection{Emission Latency Metrics}

To evaluate character-level recognition latency, we calculate each token's latency by subtracting the corresponding token's ground-truth end time from the model output timestamp. For chunk-based streaming models, each token's timestamp corresponds to the end time of the chunk. During inference, the model records the timestamps of all output tokens, while the character boundaries in the audio signals are extracted using the Forced Aligner tool as ground truth. Motivated by \cite{chang2020low, shangguan2021dissecting}, we measure two key metrics: (1) First Token Delay (FTD), the emission latency of the first token in each utterance, and (2) Last Token Delay (LTD), the emission latency of the last token in each utterance. To mitigate the impact of outliers, we report the 50th percentile (P50) and the 90th percentile (P90) of all utterances, excluding abnormal sentences. The estimated timestamps are obtained using a GMM/HMM model trained with the Kaldi toolkit \cite{ravanelli2019pytorch}.

\begin{figure}[t]
  \centering
  \includegraphics[width=\linewidth]{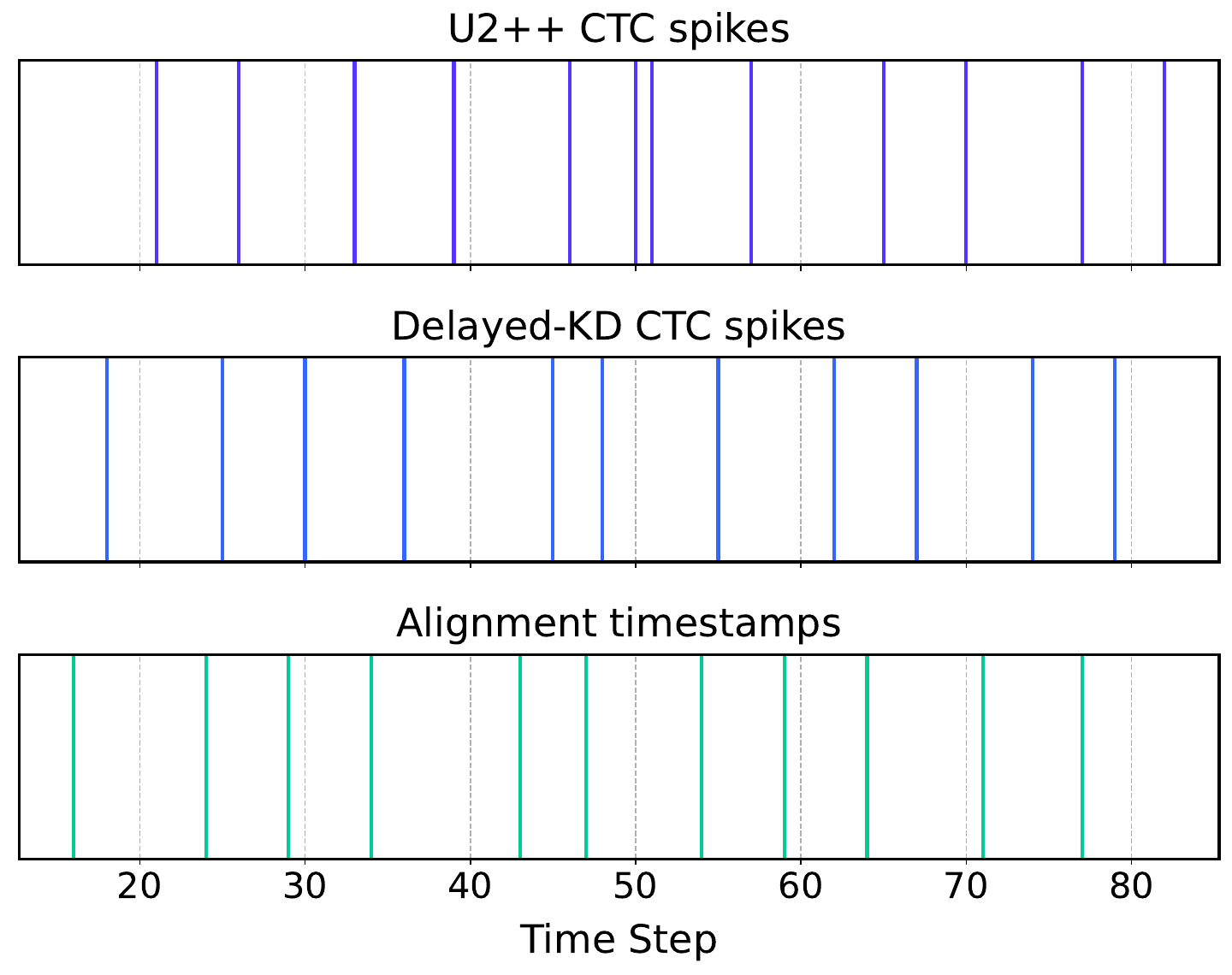}
  \caption{Comparison of CTC spike distributions between U2++ and Delayed-KD. Colored lines represent CTC spikes, and dashed lines align time axis positions.}
  \label{fig:ctc_spike_distributions}
\end{figure}

\subsection{Results on ASIHELL-1}

Table~\ref{tab:model_comparison} presents the results on AISHELL-1, including latency, CER, and token emission delay, specifically FTD and LTD. Additionally, for Delayed-KD, we compare the impact of different Temporal Alignment Buffer (TAB) settings on CER and token emission delay.

\begin{table*}[ht]
    \centering
    \caption{Comparison of CER, FTD, and LTD results for our proposed Delayed-KD with different Temporal Alignment Buffer (TAB) settings and other baseline models on AISHELL-1 test set. TAB value of 0 represents frame-by-frame distillation. Streaming refers to frame-level streaming decoding while rescoring represents an utterance-level re-scoring decoding.}
    \label{tab:model_comparison}
    \renewcommand{\arraystretch}{1.1} % 调整行距
    \setlength{\tabcolsep}{3pt} % 缩小列间距
    \begin{tabular}{c c c c cc cc cc}
        \hline
        \multirow{2}{*}{\textbf{Model}} & \textbf{Temporal Alignment} & \textbf{Decoding} & \textbf{Latency} & \multicolumn{2}{c}{\textbf{CER (\%)}} & \multicolumn{2}{c}{\textbf{FTD (ms)}} & \multicolumn{2}{c}{\textbf{LTD (ms)}} \\
                                        & \textbf{Buffer (TAB)}       & \textbf{chunk size} &    \textbf{(ms)}                                    & \textbf{Streaming}  & \textbf{Rescoring}  & \textbf{P50}      & \textbf{P90}      & \textbf{P50}      & \textbf{P90}      \\
        \hline
        \hline
        \multirow{2}{*}{\centering U2++~\cite{wu2021u2++}}  
            & \multirow{2}{*}{-} & 8 & 320 & 6.53 & 5.44 & 360 & 410 & 310 & 340 \\
            &  & 1 & 40  & 7.76 & 5.98 & 170 & 210 & 120 & 150 \\
        \cdashline{1-10}
        %\hline
        Fast-U2++~\cite{liang2023fast} & - & 4/24 & 160/960  & 7.34 & \textbf{5.06} & 170 & 220 & 70 & 120 \\
        \cdashline{1-10}
        CUSIDE-T~\cite{zhao2024cuside} & - & 10 & 400  & \textbf{6.02} & 5.51 & - & - & - & - \\
        \hline
        \multirow{5}{*}{\centering Delayed-KD} 
            & 0 ms             & \multirow{5}{*}{1} & \multirow{5}{*}{40} & 7.02          & 5.77          & \textbf{100} & \textbf{140} & \textbf{50} & \textbf{90} \\
            & 40 ms   &                   &                     & 6.67          & 5.65          & 120          & 160          & 70          & 110          \\
            & 80 ms   &                   &                     & 6.11 & 5.42 & 140          & 180          & 90          & 130          \\
            & 120 ms  &                   &                     & 6.14          & 5.47          & 160          & 200          & 110          & 150          \\
            & 160 ms  &                   &                     & 6.15          & 5.48          & 180          & 220          & 130          & 160          \\
        \hline
    \end{tabular}
\end{table*}

\textbf{Performance superiority of Delayed-KD.} Delayed-KD achieves an enhanced optimization of accuracy and latency on AISHELL-1 through its TAB mechanism and delayed knowledge distillation approach. As shown in Table~\ref{tab:model_comparison}, in rescoring decoding mode, Delayed-KD achieves a CER of 5.42\% at 40 ms latency, even comparable to another competitive solution U2++ running at 320 ms latency. While at the same 40 ms latency, Delayed-KD delivers a 9.34\% relative CER reduction. In streaming decoding mode, Delayed-KD demonstrates enhanced frame-level decoding capabilities, achieving a 21.26\% relative CER reduction over U2++ at 40 ms latency and even surpassing U2++'s performance at 320 ms latency. Through dynamic TAB size adjustment, Delayed-KD significantly reduces both FTD and LTD compared to U2++ across 320 ms and 40 ms latency settings, effectively minimizing token emission delay. As visually illustrated in Figure~\ref{fig:ctc_spike_distributions}, the CTC spikes of Delayed-KD occur notably earlier than those of U2++, providing clear evidence of its improved alignment and reduced token emission latency.

Delayed-KD demonstrates superior low-latency and high-accuracy capabilities compared to other advanced streaming models. Fast-U2++ uses a hybrid block strategy (4/24 for streaming/rescoring), while CUSIDE-T operates at a minimum latency of 400 ms. In rescoring decoding mode, Delayed-KD at 40 ms latency outperforms CUSIDE-T at 400 ms latency, while its accuracy is lower than that of Fast-U2++, which is expected given the latter's 24 times higher latency. In streaming decoding mode, Delayed-KD at 40 ms surpasses Fast-U2++ at 160 ms and approaches CUSIDE-T's performance, demonstrating its ability to balance latency and accuracy effectively.

\textbf{Analysis of TAB Settings and Trade-offs.} We also investigate the impact of different TAB sizes on CER and token emission latency. As shown in Table~\ref{tab:model_comparison}, direct frame-to-frame distillation achieves the lowest token emission latency but suboptimal CER. Increasing the TAB size improves CER significantly, with the 80 ms TAB setting yielding the best results: rescoring CER of 5.42\% and streaming CER of 6.11\%. This corresponds to a relative CER reduction of 6.07\% and 12.96\%, respectively, compared to frame-to-frame distillation. Although larger TAB sizes increase emission latency, including FTD and LTD, Delayed-KD maintains accuracy advantages while keeping latency lower than U2++, effectively balancing accuracy and latency.

\begin{table}[h!]
    \caption{Streaming ASR results in streaming and rescoring modes across varying distillation weight ($\alpha$)}
    \label{table:distillation_weights}
    \centering
    \renewcommand{\arraystretch}{1.1} % 调整行距，减少表格高度
    \setlength{\tabcolsep}{4pt} % 缩小列间距，减少宽度
    \begin{tabular}{c c c}
        \hline
        \multirow{2}{*}{\textbf{Distillation Weight ($\alpha$)}} & \multicolumn{2}{c}{\textbf{CER (\%)}} \\ 
        \cline{2-3}
        & \textbf{Streaming} & \textbf{Rescoring} \\ 
        \hline
        4   & 6.54 & 5.62 \\
        20  & 6.35 & 5.56 \\
        100 & \textbf{6.11} & \textbf{5.42} \\
        200 & 7.10 & 5.87 \\
        400 & 7.99 & 6.30 \\
        \hline
    \end{tabular}
\end{table}

\textbf{Analysis of Distillation Weight Configurations.} The distillation weight plays a critical role in joint training, balancing delayed distillation loss and total loss. As shown in Table~\ref{table:distillation_weights}, too low a weight weakens the effectiveness of knowledge distillation, while weights exceeding 100 degrade performance in both Streaming and rescoring modes. We hypothesize that this decline is primarily due to overfitting to the non-streaming CTC posterior probability distributions, negatively impacting generalization.

\subsection{Results on WenetSpeech}

We further conduct a comparative evaluation of Delayed-KD and U2++ on the large-scale WenetSpeech dataset. Using the optimal hyper-parameter settings from the AISHELL-1 experiments, we perform experiments on the \textit{Test\_Meeting} and \textit{Test\_Net} test sets of WenetSpeech. As shown in Table~\ref{tab:wenetspeech_results}, consistent with the conclusions from the AISHELL-1 experiments, Delayed-KD achieves comparable results at 40 ms latency to U2++ at 320 ms latency, and performs significantly better under the same latency conditions. One expected observation is that Delayed-KD is particularly effective in improving the accuracy of the streaming mode, yielding better results than the rescoring mode. These results indicate that Delayed-KD maintains robust performance even on large-scale datasets.

\begin{table}[h!]
    \centering
    \caption{Streaming ASR results on WenetSpeech test set (\textit{Test\_Meeting}, \textit{Test\_Net}). CER results are reported as streaming / rescoring.}
    \renewcommand{\arraystretch}{1.2}
    \setlength{\tabcolsep}{2pt}
    \resizebox{\linewidth}{!}{
        \begin{tabular}{ c c c c c c }
            \hline
            \multirow{2}{*}{\textbf{Model}} & \textbf{Decoding} & \textbf{Latency} & \multicolumn{2}{c}{\textbf{CER(\%)}} \\
            \cline{4 - 5}
            & \textbf{{chunk size}} & \textbf{(ms)} & \textbf{Test\_Meeting} & \textbf{Test\_Net} \\
            \hline
            \multirow{2}{*}{U2++} & 8 & 320 & 17.80 / \textbf{16.76} & 11.54 / \textbf{12.23} \\
            & 1 & 40 & 25.32 / 23.52 & 16.51 / 19.06 \\
            \hline
            Delayed-KD & 1 & 40 & \textbf{17.53} / 16.96 & \textbf{11.37} / 12.75 \\
            \hline
        \end{tabular}
    }
    \label{tab:wenetspeech_results}
\end{table}

\section{Conclusion}

In this paper, we propose Delayed-KD, a novel delayed knowledge distillation method that distills the CTC posterior probabilities from a non-streaming teacher model. By introducing a Temporal Alignment Buffer (TAB) during training, Delayed-KD aligns the CTC outputs of streaming and non-streaming models, effectively reducing mismatches in the distillation process. Experimental results on the AISHELL-1 and WenetSpeech datasets demonstrate that our method achieves an optimal balance between latency and accuracy.

\bibliographystyle{IEEEtran}
\bibliography{mybib}

\end{document}